\documentclass[]{article}
\usepackage{hyperref}       
\usepackage{url}   
\usepackage{hyperref}       
\usepackage{url}            
\usepackage{booktabs}       
\usepackage{amsfonts}       
\usepackage{nicefrac}       
\usepackage{microtype}      
\usepackage{xcolor}         
\usepackage{amsmath,amssymb,mathtools,bm}
\usepackage{siunitx}
\usepackage{subcaption} 

\providecommand{\tightlist}{%
	\setlength{\itemsep}{0pt}\setlength{\parskip}{0pt}}
\usepackage[parfill]{parskip}

\title{\texttt{LenslessPiCam}: A Hardware and Software Platform for Lensless Computational Imaging with a Raspberry Pi}
\author{
	Eric Bezzam\thanks{Corresponding author: \texttt{eric.bezzam@epfl.ch}}
	\and
	Sepand Kashani
	\and 
	Martin Vetterli
	\and
	Matthieu Simeoni
}
\date{}

\begin{document}

\maketitle

\section*{Summary}

Lensless imaging seeks to replace/remove the lens in a conventional imaging system. The earliest cameras were in fact 
lensless, relying on 
long exposure times to form images on the other end of a small aperture in a 
darkened room/container (\textit{camera obscura}). The introduction of a lens
allowed for more light throughput and therefore shorter exposure 
times, while retaining sharp focus. The 
incorporation of digital sensors
readily enabled the use of computational imaging techniques to post-process and enhance raw images (e.g.\ via deblurring, inpainting, denoising, sharpening).
Recently, imaging scientists have started leveraging computational imaging as an integral part of lensless imaging systems, allowing them to form viewable images from the highly multiplexed raw measurements of lensless cameras (see~\cite{boominathan2022recent} and references therein for a comprehensive treatment of lensless imaging).
This represents a real paradigm shift in camera system design as there is more flexibility to cater the hardware to the application at hand (e.g.\ lightweight or flat designs).
This increased flexibility comes however at the price of a more demanding post-processing of the raw digital recordings and a tighter integration of sensing and computation, often difficult to achieve in practice due to inefficient interactions between the various communities of scientists involved. 
With \texttt{LenslessPiCam}, we provide an easily accessible hardware and software 
framework to enable researchers, hobbyists, and students to implement and 
explore practical and computational aspects of lensless imaging. We also provide
detailed guides and exercises so that \texttt{LenslessPiCam} can be used as an educational 
resource, and point to results from our graduate-level signal processing course.\\

\noindent GitHub repo: \url{https://github.com/LCAV/LenslessPiCam}

\section*{Statement of need}

\begin{figure}[t!]
	\centering
	\begin{subfigure}{.48\textwidth}
		\centering
		\includegraphics[width=0.99\linewidth]{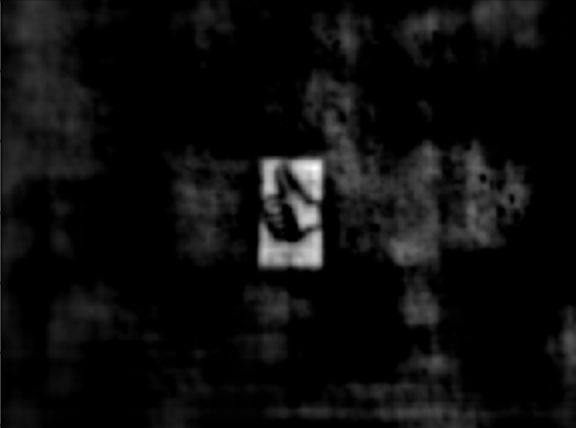}
		\caption{}
		\label{fig:diffusercam}
	\end{subfigure}
	\begin{subfigure}{.48\textwidth}
		\centering
		\includegraphics[width=0.99\linewidth]{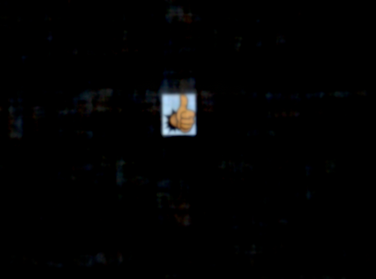} 
		\caption{}
		\label{fig:lenslesspicam}
	\end{subfigure}
	\caption{ADMM reconstruction of thumbs-up on a phone 40 cm away.}
	\label{fig:compare_cams}
\end{figure}

Being at the interface of hardware, software, and algorithm design, the field of
lensless imaging necessitates a broad array of competences that might deter 
newcomers to the field. The purpose of \texttt{LenslessPiCam} is to provide a
complete toolkit with cheap, 
accessible hardware designs and open-source software, that also achieves 
satisfactory results in order to explore novel ideas for hardware, software, and
algorithm design.

The DiffuserCam tutorial~\cite{diffusercam} served as a great starting point to the present toolkit as it demonstrates that a working lensless camera can be
built with cheap hardware: a Raspberry Pi, the Camera Module 2,\footnote{\url{www.raspberrypi.com/products/camera-module-v2}} and a piece 
of tape. The
authors also provide Python implementations of two image reconstruction 
algorithms:
\begin{enumerate}
	\tightlist
	\item Variants of gradient descent (GD) with a non-negativity constraint,
	\item The alternating direction method of multipliers (ADMM)~\cite{boyd2011distributed}
	with an additional total variation (TV) prior.
\end{enumerate}
Moreover,
detailed guides explain how to build their camera and give intuition behind the reconstruction algorithms. 

Unfortunately, the resolution of the reconstructed images is poor (see Figure~\ref{fig:diffusercam}) and the processing pipeline is limited to
grayscale reconstruction. With \texttt{LenslessPiCam}, we improve the resolution by
using the newer HQ camera\footnote{\url{www.raspberrypi.com/products/raspberry-pi-high-quality-camera/}} as well as a more versatile and generic RGB computational imaging pipeline. The latter is built upon the Python library Pycsou~\cite{pycsou}, a universal and reusable software environment providing key computational imaging functionalities and tools with great
modularity and interoperability. This results in a more flexible and accurate reconstruction workflow, allowing for the quick prototyping of advanced post-processing schemes with more sophisticated image priors.
See Figure~\ref{fig:lenslesspicam} for an example image obtained with our lensless imaging framework.

\texttt{LenslessPiCam} is designed to be used by researchers, hobbyists, and students.
In the past, we have found such open-source hardware and software platforms to be a valuable 
resource for researchers~\cite{bezzam2017hardware} and students alike~\cite{bezzam2019teaching}.
	
\section*{Contributions}


\begin{figure}[t!]
	\centering
	\begin{subfigure}{.3\textwidth}
		\centering
		\includegraphics[width=0.99\linewidth]{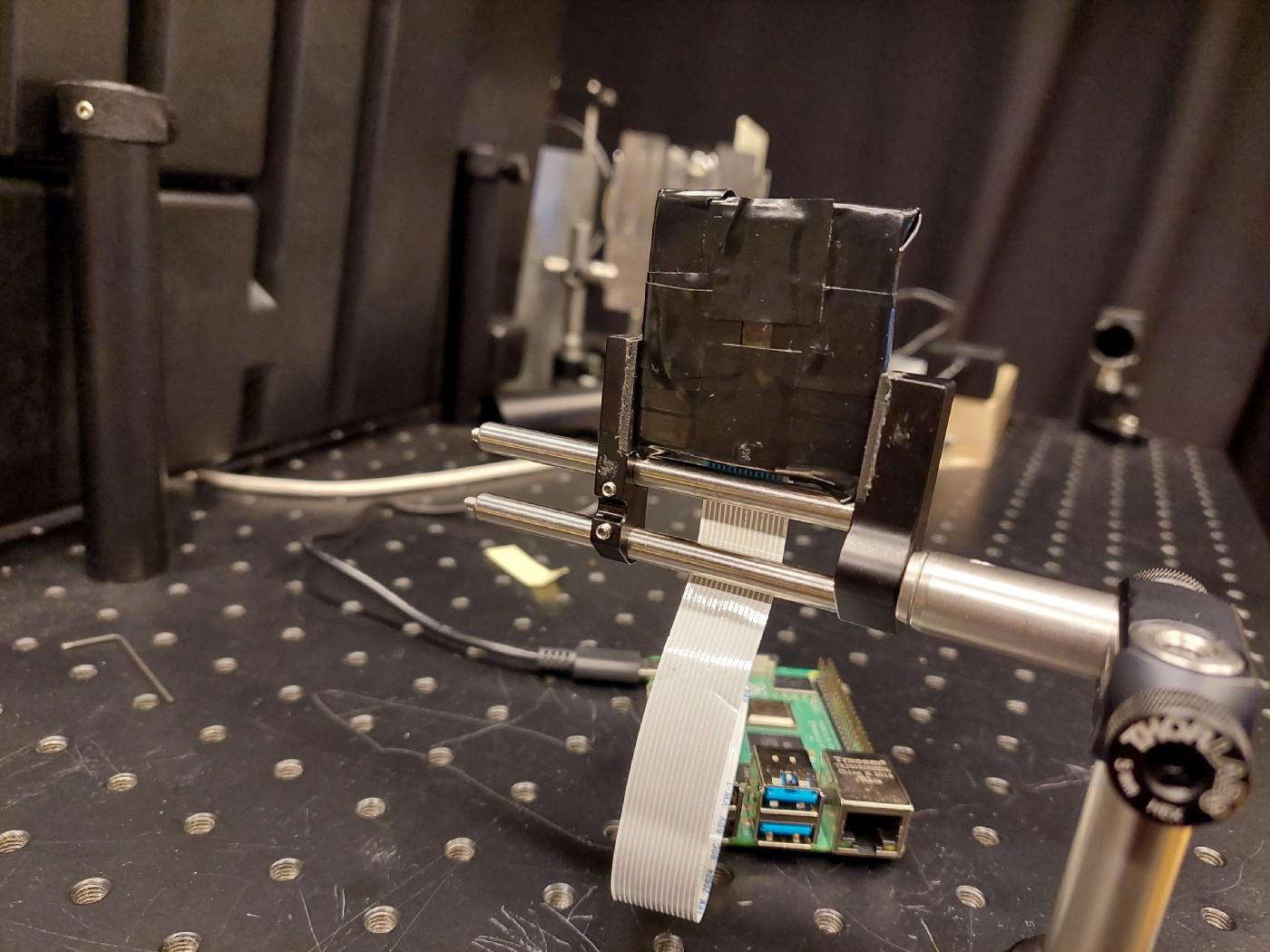}
		\caption{}
		\label{fig:camera}
	\end{subfigure}
	\begin{subfigure}{.3\textwidth}
		\centering
		\includegraphics[width=0.99\linewidth]{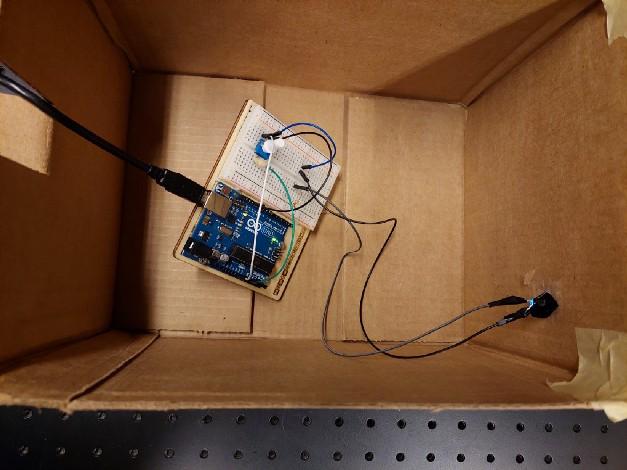} 
		\caption{}
		\label{fig:point_source_inside}
	\end{subfigure}
	\begin{subfigure}{.3\textwidth}
		\centering
		\includegraphics[width=0.99\linewidth]{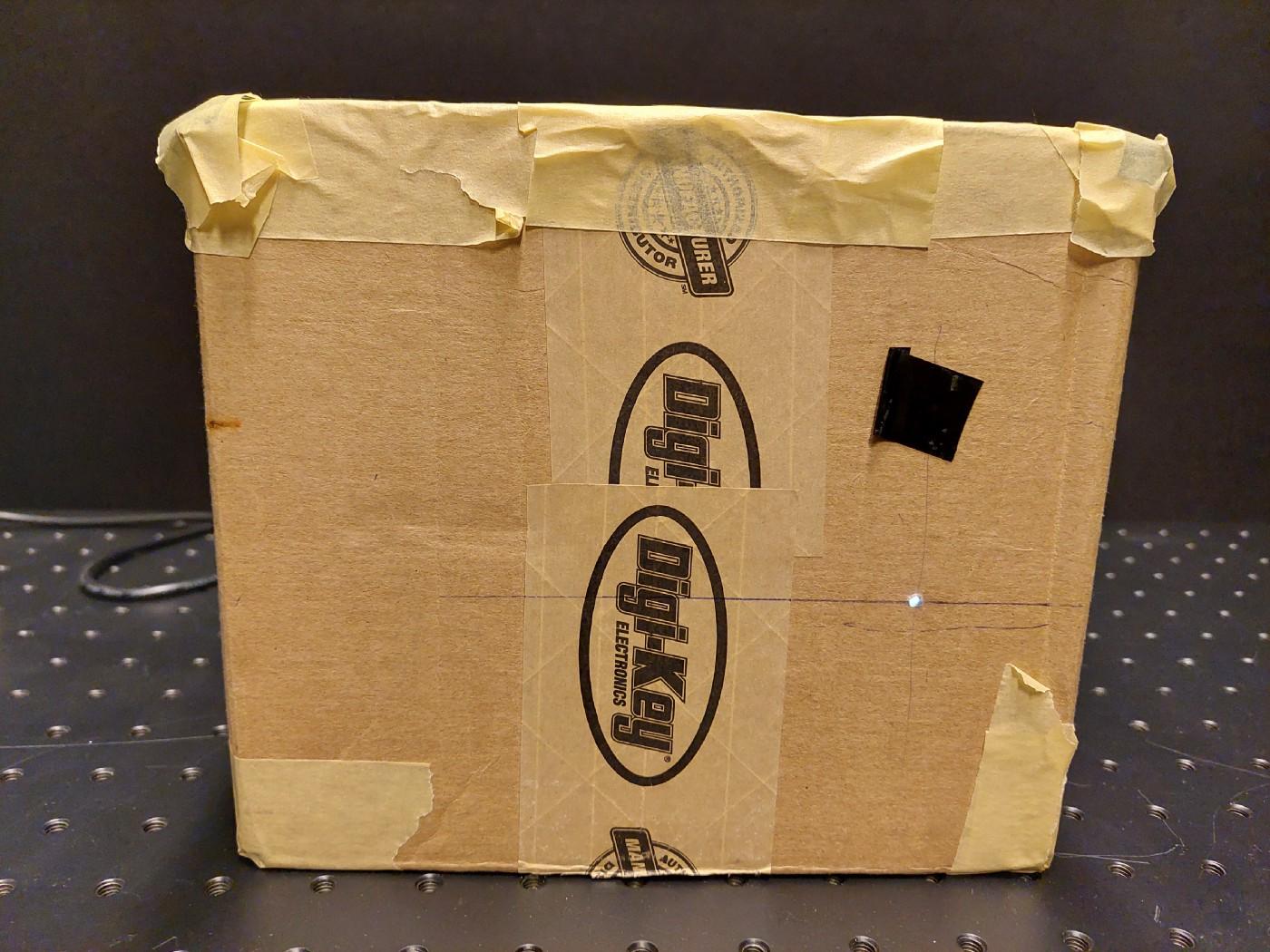} 
		\caption{}
		\label{fig:point_source}
	\end{subfigure}
	\caption{(a) LenslessPiCam, (b) point source generator (inside) and (c) (outside).}
	\label{fig:hardware}
\end{figure}

With respect to the DiffuserCam tutorial~\cite{diffusercam}, we have made the following
contributions. In terms of hardware, as shown in Figure~\ref{fig:hardware}, we:
\begin{itemize}
	\tightlist
	\item make use of the HQ camera sensor ($\$50$): 4056$\times$3040 pixels (12.3 MP) and \SI{7.9}{\milli\meter} sensor diagonal, compared to 3280$ \times $2464 pixels (8.1 MP) and \SI{4.6}{\milli\meter} sensor diagonal for the Camera Module 2 ($\$30$), 
	\item provide the design and firmware for a cheap point source generator (needed 
	for calibration), which consists of an Arduino, a white LED, and a cardboard box.
\end{itemize}

\noindent With respect to reconstruction algorithms, we:
\begin{itemize}
	\tightlist
	\item provide significantly faster implementations of GD and ADMM, i.e. around 3$ \times $ 
	reduction in computation time,
	\item extend the above reconstructions to RGB,
	\item provide an object-oriented structure that is easy to extend for exploring new 
	algorithms,
	\item provide an object-oriented interface to Pycsou for solving
	lensless imaging inverse problems. Pycsou is a Python package
	for solving inverse problems of the form
	\begin{equation}\label{eq:fourier}
		\min_{\mathbf{\alpha}\in\mathbb{R}^N} \,F(\mathbf{y}, \mathbf{G} \mathbf{\alpha})\quad+\quad \lambda\mathcal{R}(\mathbf{\alpha}),
	\end{equation}
	where $F$ is a data-fidelity term between the observed and predicted 
	measurements $\mathbf{y}$ and $\mathbf{G}\mathbf{\alpha}$ respectively, 
	$\mathcal{R}$ is a regularization component (could consist of more than one prior), 
	and $\lambda >0$ controls the amount of regularization.
\end{itemize}
	
\noindent We also provide functionality to:
\begin{itemize}
	\tightlist
	\item remotely capture Bayer data with the proposed camera,
	\item convert Bayer data to RGB or grayscale,
	\item quantitatively evaluate the point spread function (PSF) of the lensless camera,
	\item remotely display data on an external monitor, which can be used to automate 
	raw data measurements to, e.g., gather a dataset,
	\item collect MNIST remotely or from the Raspberry Pi,
	\item evaluate reconstructions on a variety of metrics: MSE, PSNR, SSIM, LPIPS~\cite{zhang2018perceptual}.
\end{itemize}
	
\noindent Finally, we have written a set of Medium articles to guide users through the 
process of building, using and/or teaching with the proposed lensless camera.\footnote{An overview of these articles can be found here: \url{https://medium.com/@bezzam/a-complete-lensless-imaging-tutorial-hardware-software-and-algorithms-8873fa81a660}}
The articles include a set of solved exercises and problems for teaching purposes (solutions available to instructors on request).
	
In the following sections, we describe some of these contributions, and quantify them (where appropriate).
	
\section*{High-level functionality}

The core algorithmic component of \texttt{LenslessPiCam} is the abstract class \\
\texttt{lensless.ReconstructionAlgorithm}. The three reconstruction strategies 
available in \texttt{LenslessPiCam} derive from this class:
\begin{itemize}
	\tightlist
	\item \texttt{lensless.GradientDescient}: projected GD 
	with a non-negativity constraint. Two accelerated approaches are also
	available:
	\begin{itemize}
		\tightlist
		\item \texttt{lensless.NesterovGradientDescent}~\cite{nesterov1983method}
		\item \texttt{lensless.FISTA}~\cite{beck2009fast}
	\end{itemize}
	\item \texttt{lensless.ADMM}: ADMM with a non-negativity constraint and a TV regularizer,
	\item \texttt{lensless.APGD}: accelerated proximal GD with Pycsou
	as a backend. Any differentiable or proximal operator can be used as long as it 
	is compatible with Pycsou, namely derives from one of 
	\texttt{DifferentiableFunctional} or \texttt{ProximableFunctional}.
\end{itemize}

\noindent One advantage of deriving from \texttt{lensless.ReconstructionAlgorithm} is that
functionality for iterating, saving, and visualization is already implemented. 
Consequently, using a reconstruction algorithm that derives from it boils down 
to three steps:
\begin{enumerate}
	\tightlist
	\item Creating an instance of the reconstruction algorithm.
	\item Setting the data.
	\item Applying the algorithm.
\end{enumerate}

\noindent For example, for ADMM (full example in \texttt{scripts/recon/admm.py}):
\begin{verbatim}
	recon = ADMM(psf)
	recon.set_data(data)
	res = recon.apply(n_iter=n_iter)
\end{verbatim}

A template for applying a reconstruction algorithm (including loading the data)
can be found in \texttt{scripts/recon/template.py}.

\section*{Efficient reconstruction}

In Table~\ref{tab:benchmark}, we compare the processing time of DiffuserCam's and 
 \texttt{LenslessPiCam}'s implementations for grayscale reconstruction of:
\begin{enumerate}
	\tightlist
	\item GD using FISTA and a non-negativity constraint,
	\item ADMM with a non-negativity constraint and a TV regularizer.
\end{enumerate}

The DiffuserCam implementations can be found here: \url{https://github.com/Waller-Lab/DiffuserCam-Tutorial}, while 
\texttt{lensless.APGD} and \texttt{lensless.ADMM} are used for \texttt{LenslessPiCam}. The 
comparison is done on a Lenovo Thinkpad P15 with 16 GB RAM and a 2.70 GHz
processor (6 cores, 12 threads), running Ubuntu 21.04.

\begin{table}[t!]
	\centering
	\begin{tabular}{|c|c|c|}
		\hline
		& GD & ADMM \\
		\hline
		DiffuserCam & \SI{215}{\second} & \SI{7.24}{\second} \\
		\hline
		\texttt{LenslessPiCam} & \SI{67.9}{\second} & \SI{2.76}{\second} \\
		\hline
	\end{tabular}
\caption{Benchmark grayscale reconstruction. 300 iterations for gradient descent (GD)
	and 5 iterations for alternating direction method of multipliers (ADMM).}
\label{tab:benchmark}
\end{table}

From Table~\ref{tab:benchmark}, we observe a 3.1$ \times $ reduction in computation time for
GD and a 2.6$ \times $ reduction for ADMM. This comes from:
\begin{itemize}
	\tightlist
	\item our object-oriented implementation of the algorithms, which allocates all the 
	necessary memory beforehand and pre-computes data-independent terms, such
	as forward operators from the point spread function (PSF),
	\item our use of the real-valued FFT, which is possible since we are working with 
	image intensities.
\end{itemize}

\begin{figure}[t!]
	\centering
	\begin{subfigure}{.49\textwidth}
		\centering
		\includegraphics[width=0.99\linewidth]{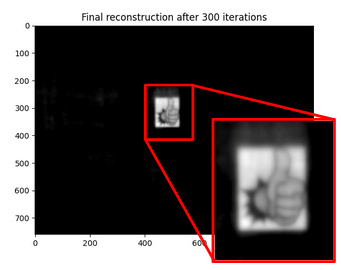}
		\caption{}
		\label{fig:gradient_descent}
	\end{subfigure}
	\begin{subfigure}{.49\textwidth}
		\centering
		\includegraphics[width=0.99\linewidth]{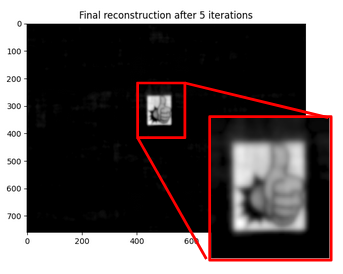} 
		\caption{}
		\label{fig:admm}
	\end{subfigure}
	\caption{Grayscale reconstruction using FISTA (a) and ADMM (b).}
	\label{fig:grayscale}
\end{figure}

Figure~\ref{fig:grayscale} shows the corresponding grayscale reconstruction for 
FISTA and ADMM, which are equivalent for both DiffuserCam and \texttt{LenslessPiCam} implementations.


\section*{Quantifying performance}

In order to methodically compare different reconstruction approaches, it is necessary to
quantify the performance. To this end, \texttt{LenslessPiCam} provides functionality
to extract regions of interest from the reconstruction and compare them with the
original image via multiple metrics:
\begin{itemize}
	\tightlist
	\item Mean-squared error (MSE),
	\item Peak signal-to-noise ratio (PSNR),
	\item Mean structural similarity (SSIM) index,
	\item Learned perceptual image patch similarity (LPIPS).
\end{itemize}

\begin{figure}[b!]
	\centering
	\includegraphics[width=0.85\linewidth]{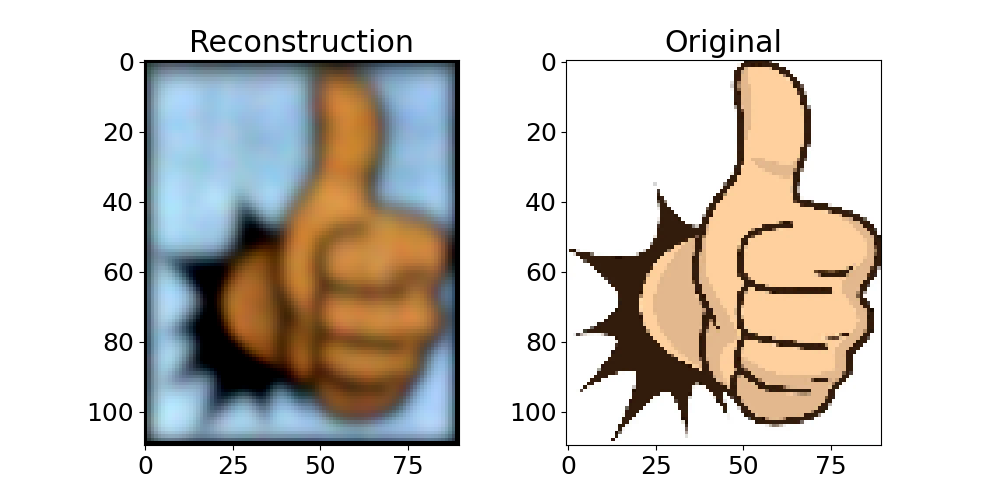} 
	\caption{Extracting region from Figure~\ref{fig:lenslesspicam} to quantify performance.}
	\label{fig:metric}
\end{figure}

Figure~\ref{fig:metric} and Table~\ref{tab:metric} is an example of how a reconstruction can be evaluated against an original
image, e.g.\ using \texttt{scripts/compute\textunderscore metrics\textunderscore original.py}

\begin{table}[b!]
	\centering
\begin{tabular}{|c|c|c|c|}
	\hline
	MSE & PSNR & SSIM & LPIPS \\
	\hline
	0.164 & 7.85 & 0.405 & 0.645 \\
	\hline
\end{tabular}
\caption{Metrics for Figure~\ref{fig:metric}.}
\label{tab:metric}
\end{table}

Sometimes it may be of interest to perform an exhaustive evaluation on a large
dataset.
While \texttt{LenslessPiCam} could be used for collecting such a dataset with the
proposed camera,\footnote{Using the remote display and capture scripts, i.e. 
	\texttt{scripts/remote\textunderscore display.py} and \texttt{scripts/remote\textunderscore capture.py} respectively.} the 
authors of~\cite{monakhova2019learned} have already collected a dataset of 25'000 
parallel measurements, namely 25'000 pairs of DiffuserCam and lensed camera images.\footnote{\url{https://waller-lab.github.io/LenslessLearning/dataset.html}}
\texttt{LenslessPiCam} offers functionality to evaluate a reconstruction algorithm on
this dataset, or a subset of it that we have prepared.\footnote{Subset of DiffuserCam Lensless Mirflickr Dataset~\cite{monakhova2019learned}
	consists of 200 files (725 MB) as opposed to 25'000 files (100 GB) of the
	original dataset. The subset can be downloaded here: \url{https://drive.switch.ch/index.php/s/vmAZzryGI8U8rcE}.} Note that this 
dataset is collected with a different lensless camera, but is nonetheless useful
for exploring reconstruction techniques.

Table~\ref{tab:Mirflickr} shows the average metric results after applying 100 iterations of ADMM
to the subset we have prepared.\footnote{Using \texttt{scripts/evaluate\textunderscore mirflickr\textunderscore admm.py}}

\begin{table}[t!]
	\centering
	\begin{tabular}{|c|c|c|c|}
		\hline
		MSE & PSNR & SSIM & LPIPS \\
		\hline
		0.0797 & 12.7 & 0.535 & 0.585 \\
		\hline
	\end{tabular}
	\caption{Average metrics for 100 iterations of ADMM on a subset (200 files) of the DiffuserCam Lensless Mirflickr Dataset.}
	\label{tab:Mirflickr}
\end{table}

\begin{figure}[t!]
	\centering
	\includegraphics[width=0.95\linewidth]{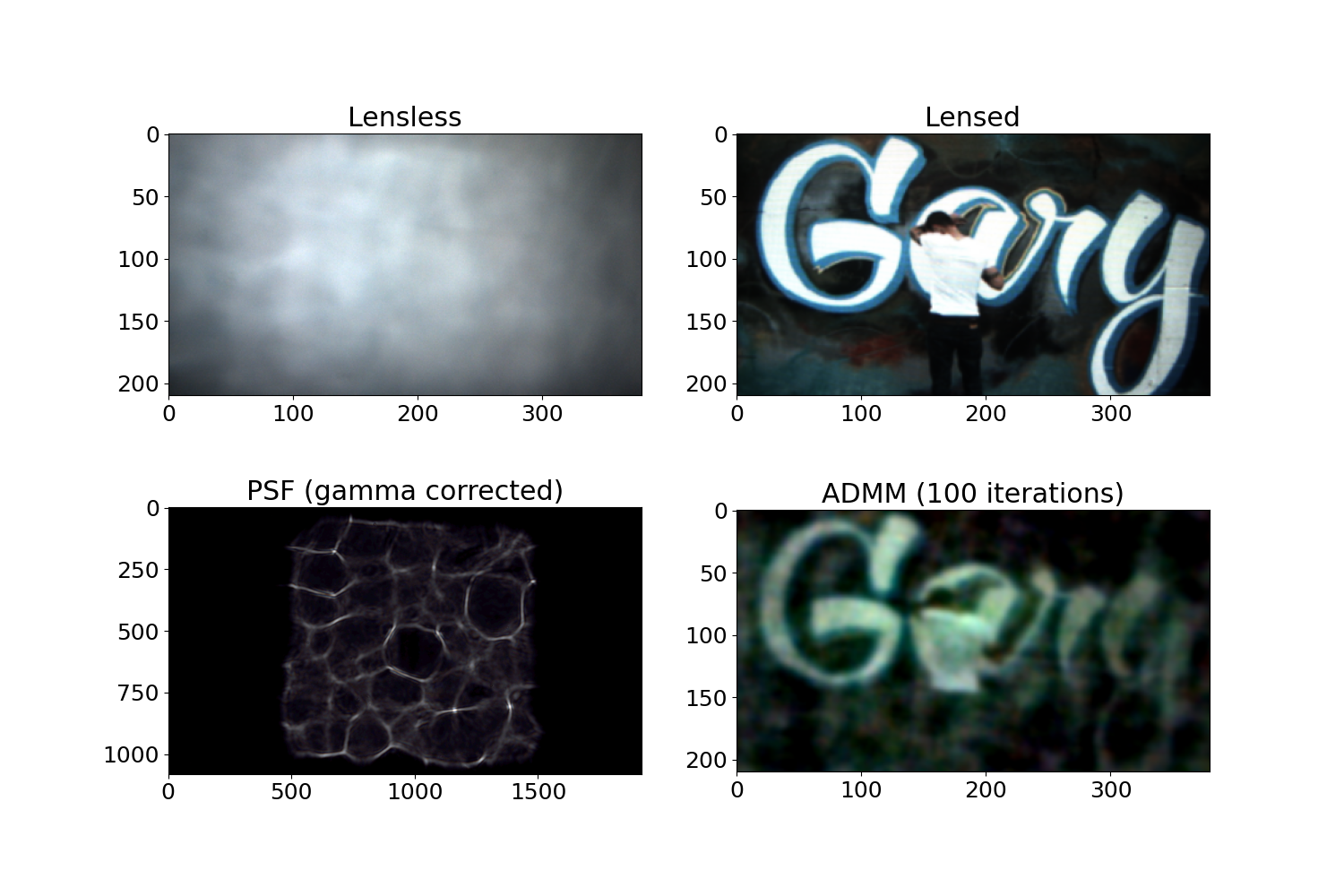} 
	\caption{Visualizing performance of ADMM (100 iterations) on a single file of the DiffuserCam Lensless Mirflickr Dataset.}
	\label{fig:dataset_single_file}
\end{figure}

\begin{table}[t!]
	\centering
	\begin{tabular}{|c|c|c|c|}
		\hline
		MSE & PSNR & SSIM & LPIPS \\
		\hline
		0.0682 & 11.7 & 0.486 & 0.504 \\
		\hline
	\end{tabular}
	\caption{Metrics for Figure~\ref{fig:dataset_single_file}.}
	\label{tab:single}
\end{table}

One can also visualize the performance on a single file of the dataset, namely
how the reconstruction changes as the number of iterations increase.\footnote{Using \texttt{scripts/apply\textunderscore admm\textunderscore single\textunderscore mirflickr.py}.} The 
final reconstruction and outputed metrics are shown in 
Figure~\ref{fig:dataset_single_file} and Table~\ref{tab:single}.

\section*{As an educational resource}

As mentioned earlier, \texttt{LenslessPiCam} can serve as an educational 
resource. We have used it in our graduate-level signal processing course for
providing experience in applying fundamental signal processing concepts and 
solving linear inverse problems. The work of our students can be found
\href{https://infoscience.epfl.ch/search?ln=en&rm=&ln=en&sf=&so=d&rg=10&c=Infoscience%2FArticle&c=Infoscience%2FBook&c=Infoscience%2FChapter&c=Infoscience%2FConference&c=Infoscience%2FDataset&c=Infoscience%2FLectures&c=Infoscience%2FPatent&c=Infoscience%2FPhysical%20objects&c=Infoscience%2FPoster&c=Infoscience%2FPresentation&c=Infoscience%2FProceedings&c=Infoscience%2FReport&c=Infoscience%2FReview&c=Infoscience%2FStandard&c=Infoscience%2FStudent&c=Infoscience%2FThesis&c=Infoscience%2FWorking%20papers&c=Media&c=Other%20doctypes&c=Work%20done%20outside%20EPFL&c=&of=hb&fct__2=LCAV&p=diffusercam}{here}.

As exercises in implementing key signal processing components, we have left some
incomplete functions in \texttt{LenslessPiCam}:
\begin{itemize}
	\tightlist
	\item \texttt{lensless.autocorr.autocorr2d}: to compute a 2D autocorrelation in the 
	frequency domain,
	\item \texttt{lensless.realfftconv.RealFFTConvolve2D}: to pre-compute the PSF's Fourier
	transform, perform a convolution in the frequency domain with the real-valued
	FFT, and vectorize operations for RGB.
\end{itemize}

We have also proposed a few reconstruction approaches to implement in \href{https://medium.com/@bezzam/lensless-imaging-with-the-raspberry-pi-and-python-diffusercam-473e47662857}{this Medium article}.

For the solutions to the above implementations, please request access to \href{https://drive.google.com/drive/folders/1Y1scM8wVfjVAo5-8Nr2VfE4b6VHeDSia?usp=sharing}{this folder} detailing the intended use. 

\section*{Conclusion}

In summary, \texttt{LenslessPiCam} provides all the necessary hardware designs and 
software to build, use, and evaluate a lensless camera with cheap and accessible
components. As we continue to use it as a research and educational platform, we
hope to investigate and incorporate:
\begin{itemize}
	\tightlist
	\item computational refocusing,
	\item video reconstruction,
	\item on-device reconstruction,
	\item programmable masks,
	\item data-driven, machine learning reconstruction techniques.
\end{itemize}

\section*{Acknowledgements}

We acknowledge feedback from Julien Fageot and the students during the first 
iteration of this project in our graduate course.

This work was in part funded by the Swiss National Science Foundation (SNSF) 
under grants CRSII5 193826 “AstroSignals - A New Window on the Universe, with 
the New Generation of Large Radio-Astronomy Facilities” (M. Simeoni), 
200 021 181 978/1 “SESAM - Sensing and Sampling: Theory and Algorithms” 
(E. Bezzam) and CRSII5 180232 “FemtoLippmann - Digital twin for multispectral 
imaging” (S. Kashani).

\bibliographystyle{plain}
\bibliography{references}

\end{document}